\documentclass[graybox]{svmult}
\usepackage{mathptmx}       
\usepackage{helvet}         
\usepackage{courier}        
\usepackage{type1cm}        
\usepackage{makeidx}         
\usepackage{graphicx}        
\usepackage{multicol}        
\usepackage[bottom]{footmisc}

\makeindex             

\def\vsin{\hbox{$v \sin i$}}  
  
\def\kms{\hbox{km\,s$^{-1}$}}  
\def\ms{\hbox{m\,s$^{-1}$}}

\def\em{\it}  
  
\def\degr{\hbox{$^\circ$}}

\def\rsun{\hbox{$R_\odot$}}

\begin{document}

\title*{Magnetic field and convection in the cool supergiant Betelgeuse}
\author{P. Petit, M. Auri\`ere, R. Konstantinova-Antova, A. Morgenthaler, G. Perrin, T. Roudier and J.-F. Donati}
\authorrunning{P. Petit et al.}
\institute{P. Petit \at IRAP, CNRS \& Universit\'e de Toulouse, \email{petit@ast.obs-mip.fr}
\and M. Auri\`ere \at IRAP, CNRS \& Universit\'e de Toulouse 
\and R. Konstantinova-Antova \at Institute of Astronomy, Bulgarian Academy of Science
\and A. Morgenthaler \at IRAP, CNRS \& Universit\'e de Toulouse 
\and G. Perrin \at Observatoire de Paris, LESIA 
\and T. Roudier \at IRAP, CNRS \& Universit\'e de Toulouse 
\and J.-F. Donati \at IRAP, CNRS \& Universit\'e de Toulouse 
}
%
%
\maketitle

\abstract{We present the outcome of a highly-sensitive search for magnetic fields\index{magnetic field} on the cool supergiant\index{cool supergiant} Betelgeuse\index{Betelgeuse}. A time-series of six circularly-polarized spectra was obtained using the NARVAL spectropolarimeter\index{spectropolarimetry} at T\'elescope Bernard Lyot (Pic du Midi Observatory), between 2010 March and April. Zeeman signatures were repeatedly detected in cross-correlation profiles, corresponding to a longitudinal component of about 1~G. The time-series unveils a smooth increase of the longitudinal field from 0.5 to 1.5~G, correlated with radial velocity fluctuations. We observe a strong asymmetry of Stokes V signatures, also varying in correlation with the radial velocity. The Stokes V line profiles are red-shifted by about 9~\kms\ with respect to the Stokes I profiles, suggesting that the observed magnetic elements may be concentrated in the sinking components of the convective flows.}

\section{Introduction}
\label{sec:intro}

The widespread signatures of magnetic activity\index{activity} in cool stars\index{star} are believed to originate from the coexistence of convection and rotation in their outer layers. Since the first dynamo\index{dynamo} models of Parker (1955), rotation is generally accepted as a central ingredient of these stellar dynamos\index{dynamo}, in a two-level action. The stellar spin is first involved in the generation of radial and latitudinal shears that are able to wind up the field lines of a seed magnetic field\index{magnetic field} around the rotation axis, resulting in the creation of a strong toroidal field component. Stellar rotation is also acting on the vertical plasma flows through the Coriolis force that succeeds at generating helical motions able to twist again the field lines of the toroidal field and generate a poloidal field component. 

If  many details of the physical processes involved in such large-scale dynamos\index{dynamo} are still a matter of debate (see e.g. Charbonneau 2010 for a review), this theoretical framework is now widely accepted to interpret the cyclical activity\index{activity} behaviour of the Sun and solar-type stars\index{star}. However, other models suggest that convection alone is able to sustain a dynamo\index{dynamo}, without any rotational effects involved (e.g. Cattaneo 1999, V\"ogler \& Sch\"ussler 2007). Turbulent dynamo\index{dynamo} action may be responsible for the smallest-scale magnetic elements observed on the solar surface (Lites et al. 2008), although intranetwork magnetic elements may also result from the decay of larger magnetic regions (created by the global dynamo\index{dynamo}) through the continuous convective mixing of the solar upper layers.

\begin{figure}[t]
\sidecaption[t]
\includegraphics[width=10cm]{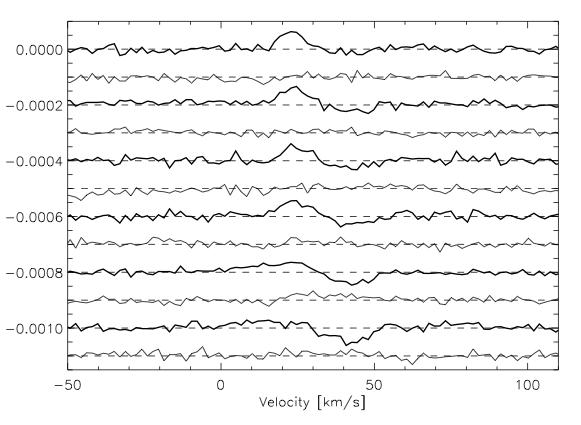}
%
%
\caption{Time-series of Stokes V (bold lines) and their corresponding ``null'' line profiles (thin line below). The successive profiles are shifted vertically for display clarity. After Auri\`ere et al. (2010).}
\label{fig:stokesv}
\end{figure}

Since both global and small-scale dynamos\index{dynamo} may be simultaneously active in the Sun, it is not easy to disentangle the respective magnetic outcome of these two different processes. A promising way to reach this goal consists in observing a star\index{star} with no rotation at all, or at least a star\index{star} rotating so slowly that the onset of a global dynamo\index{dynamo} in its internal layers is unlikely. If stellar spectropolarimetry\index{spectropolarimetry} is our best asset to detect a magnetic field\index{magnetic field} in a non-rotating star\index{star}, the polarimetric detection of Zeeman signatures is mostly insensitive to small-scale magnetic elements as those expected to be generated by a local dynamo\index{dynamo}. This issue is unescapable for solar-type dwarfs, on which millions of photospheric convective cells are visible at any time, resulting in a highly tangled intranetwork field pattern. Cool supergiant\index{cool supergiant} stars\index{star} may offer a rare opportunity to circumvent this problem, since their convective cells are expected to be much larger than on the Sun, with only a few of them covering the stellar surface (Schwarzschild 1975, Chiavassa et al. 2010), so that the spatial scale of convection may be sufficiently large to limit the mutual cancellation of Zeeman signatures of close-by magnetic elements with opposite polarities.

To test the feasability of magnetic field\index{magnetic field} detection in cool supergiant\index{cool supergiant} stars\index{star}, we have concentrated on Betelgeuse\index{Betelgeuse} ($\alpha$ Orionis), one of the brightest members of this class. We briefly summarize a few well-known properties of this object, present the spectropolarimetric\index{spectropolarimetry} observations gathered with NARVAL and discuss the first outcome of this project. 

\section{The cool supergiant\index{cool supergiant} $\alpha$ Orionis}

With a M2Iab spectral type classification, Betelgeuse\index{Betelgeuse} can be taken as the  prototype of cool supergiant\index{cool supergiant} stars\index{star}. Thanks to its proximity to the Earth, it was the first star\index{star} to have its radius determined using interferometry (Michelson \& Pease 1921), and a recent value of $645 \pm 129$\rsun\ was derived in near-infrared interferometry (Perrin et al. 2004). A significant scatter is observed in radius estimates depending on the adopted wavelength domain, because of the extension and thermal structure of the stellar atmosphere (see e.g. Uitenbroek et al. 1998, and references therein). The efficient mass-loss of Betelgeuse\index{Betelgeuse} is at the origin of an inhomogeneous distribution of gas and dust, detected up to a few tens of stellar radii (Kervella et al. 2011). 

Using spatially-resolved, high-resolution UV spectroscopy with the HST, Uitenbroek et al. (1998) were able to propose a rotation period of about 17~yr and a low inclination of the rotation axis, of about 20\degr. The basic ingredients of a global dynamo\index{dynamo}, primarily based on rotation, are therefore likely missing in $\alpha$ Orionis. 

\section{Surface magnetic field\index{magnetic field}}

\begin{figure}[t]
\sidecaption[t]
\includegraphics[width=10cm]{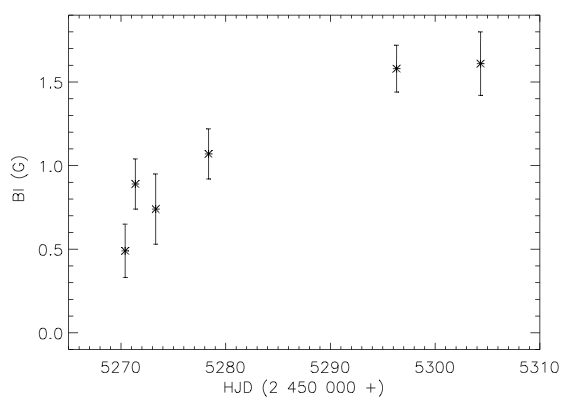}
%
%
\caption{Estimates of the disc-integrated, line-of-sight component of the magnetic field\index{magnetic field}, from 2010 March 14 to 21010 April 17 (Auri\`ere et al. 2010).}
\label{fig:bl}       
\end{figure}

Spectropolarimetric\index{spectropolarimetry} observations of Betelgeuse\index{Betelgeuse} were obtained using the NARVAL spectropolarimeter\index{spectropolarimetry} (Auri\`ere 2003), installed at T\'elescope Bernard Lyot\footnote{The Bernard Lyot Telescope is operated by the Institut National des Sciences de l'Univers of the Centre National de la Recherche Scientifique of France.}, Pic du Midi Observatory (France), in a highly-sensitive search for a weak surface magnetic field\index{magnetic field} (Auri\`ere et al. 2010). The data were collected during 6 nights in 2010, from March 14 to April 17. They consist of high-resolution ($R=65,000$) echelle spectra in light intensity (Stokes I) and circular polarization (Stokes V), offering an almost contiguous coverage of the wavelength interval between 370~nm and 1,000~nm.  Each Stokes V spectrum is built from a sequence of 4 exposures taken with different azimuths of the polarimetric optics. Using a different combination of the sub-exposures, a ``null'' control spectrum is also computed, which should not contain any detectable signature. To reach a high signal-to-noise ratio and avoid any saturation of the detector, a sequence of 16 to 20 spectra was acquired for every individual night and later averaged.

To lower further the noise level and improve the detectability of Zeeman signatures, all spectra were processed using the Least-Squares-Deconvolution (LSD) cross-correlation technique (Donati et al. 1997, Kochukhov et al. 2010). By doing so, mean line profiles were computed, using a list of 15,000 photospheric atomic lines corresponding to the atmospheric parameters of Betelgeuse\index{Betelgeuse}. The resulting time-series of Stokes V profiles is plotted in Fig. \ref{fig:stokesv}. Circularly polarized signatures, most likely generated through the Zeeman effect, are detected above noise level at the radial velocity of the star\index{star}. 

Using the centre-of-gravity technique (Rees \& Semel 1979), the Stokes V signatures can be translated into estimates of the line-of-sight component ($B_l$) of the magnetic field\index{magnetic field} (Auri\`ere et al 2010). The series of $B_l$ measurements are plotted in Fig. \ref{fig:bl}. They display an average value of the order of 1~G. If the variability of Stokes V profiles is barely visible to the naked eye, $B_l$ estimates enable one to unveil a first type of variability of the Zeeman signatures, with a regular increase of the field strength over our observing window, from 0.5 to 1.5~G.  

\section{Surface convection and magnetic activity\index{activity}}

\begin{figure}[t]
\sidecaption[t]
\includegraphics[width=14cm]{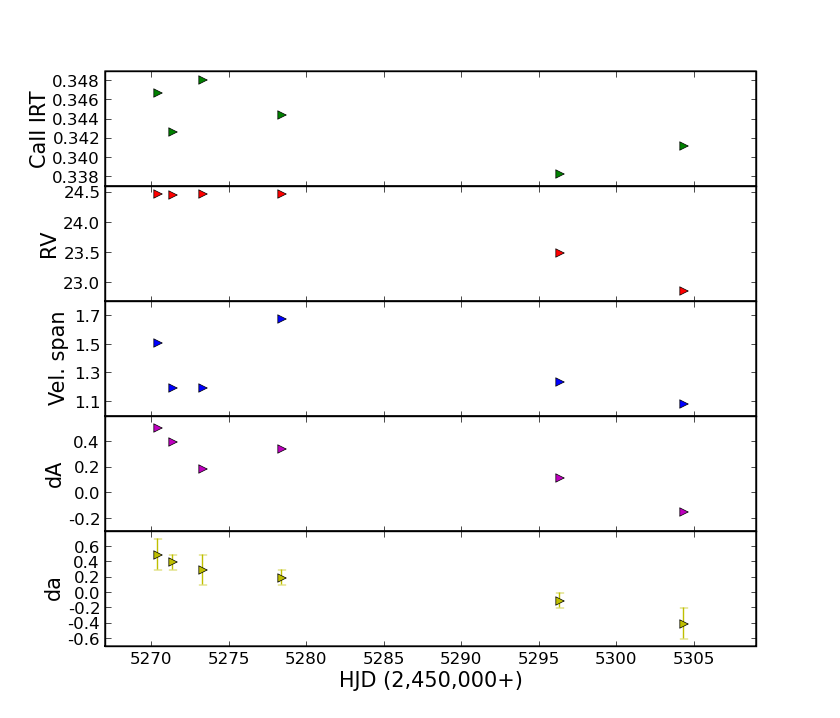}
%
%
\caption{Additional activity\index{activity} indicators derived from NARVAL spectra. From top to bottom, we plot the Ca II IRT index, the mean radial velocity of Stokes I LSD profiles (RV) in \kms, the velocity span of bisectors calculated from Stokes I LSD profiles (\kms), the relative {\em surface} asymmetry (dA) and the relative {\em amplitude} asymmetry (da) of Stokes V LSD profiles. Statistical error bars are not indicated in the plots whenever they are smaller than the symbol size.}
\label{fig:activity}       
\end{figure}

The wide spectral coverage and high spectral resolution of NARVAL spectra give access to a number of classical activity\index{activity} indicators that provide us with a useful set of additional measurements, carrying information that complement the longitudinal field estimates.

\subsection{Chromospheric emission}

The stellar chromospheric flux can be inferred from the various chromospheric lines showing up in NARVAL intensity spectra. We choose here to concentrate on the Ca II infrared triplet, located in a spectral domain where the signal-to-noise ratio is generally high in Betelgeuse\index{Betelgeuse} spectra. 

An emission index is constructed by integrating the fluxes $Ca_1$,  $Ca_2$ and $Ca_3$ in three rectangular bandpasses, centered around the three component of the Ca II triplet (at 849.8, 854.2 and 866.2 nm), each one with a width of 0.2 nm. We also integrate the flux in two rectangular bands $C_1$ and $C_2$ with a width of 0.5~nm in the neighbouring spectral domain (centered around 847.58 and 870.49 nm) to normalize the index, which is therefore computed as follows:

\begin{equation}
I = \frac{Ca_1 + Ca_2 + Ca_3}{C_1 + C_2} 
\end{equation}

\noindent The temporal evolution of this index is plotted in Fig. \ref{fig:activity}. Significant fluctuations are observed. If a global decrease is visible between March and April, the evolution is not as smooth as the increase of $B_l$. The Ca II emission is generally expected to be correlated to the local magnetic field\index{magnetic field} strength, so that the (loose) anti-correlation between $B_l$ and the Ca II index might seem paradoxical. We stress, however, that in the case of stars\index{star} with low \vsin\ values, the polarimetric signatures are specifically sensitive to the low-order component of the magnetic topology, while the chromospheric emission retains information about a wide range of spatial scales (Petit et al. 2008). The different temporal trend of both tracers is therefore not a surprise in the context of a magnetic field\index{magnetic field} probably shaped by a complex pattern of convective flows.   

\subsection{Radial velocity and profile bisectors}

The radial velocities ($RV$) of Stokes I LSD profiles, derived from the fitting of a gaussian on the line profiles, are plotted in Fig. \ref{fig:activity}, following previous work by Gray (2008). The wavelength calibration is performed using telluric lines recorded in NARVAL spectra. For solar-type dwarfs, the resulting $RV$ stability is of the order of 15 to 30 \ms\ (Moutou et al. 2007). A similar accuracy can be expected for Betelgeuse\index{Betelgeuse}, for which LSD profiles benefit from a much larger number of spectral lines. The observed fluctuations are far above this limit, reaching about 1.5 \kms. The evolution is smooth and anti-correlated to $B_l$ measurements.

The Stokes I LSD profiles of Betelgeuse\index{Betelgeuse} are also highly asymmetric, with a red wing of the line much wider than the blue wing (Auri\`ere et al. 2010). Bisectors computed from the LSD profiles confirm this trend, with a clear deviation towards the red near the continuum. The total velocity span of the bisectors (plotted in Fig. \ref{fig:activity}) is comprised between 1 and 2 \kms. Fluctuations are less organized than those recorded in $RV$, with a marginal trend to display a lower bisector span in April. 

\subsection{Stokes V asymmetry and zero-crossing of Stokes V profiles}

The double-peaked Stokes V profiles of Betelgeuse\index{Betelgeuse} all possess a significant level of asymmetry, in the sense that the blue and red lobes of the profiles have a different amplitude and delimit a different surface. This property of Stokes V signatures is widely observed on the Sun (Viticchi\'e \& Sanchez Almeida 2011) and is generally linked to vertical gradients of both velocity and magnetic field\index{magnetic field} in magnetic elements (L\'opez Ariste 2002).

To quantify this observation, we use the approach of Petit et al. (2005) and call $a_b$ and $a_r$ the amplitude of the blue and red lobes of the profile. We then derive a relative {\em amplitude} asymmetry $da = (a_b - a_r)/(a_b + a_r)$. In a similar manner, we define a relative {\em surface} asymmetry $dA = (A_b - A_r)/(A_b + A_r)$, where $A_b$ and $A_r$ are the areas of the blue and red lobe, respectively. The successive values of $dA$ and $da$ are plotted in Fig. \ref{fig:activity}. We observe a global decrease of both parameters, with a more regular trend in $da$. Both parameters change sign during the time-series.

The double-lobed shape of Stokes V profiles implies a sign reversal close to the line-core. The zero-crossing radial velocity observed for the Stokes V profiles of Betelgeuse\index{Betelgeuse} is of about 33~\kms\ (with no clear temporal trend), while the core of the Stokes I profile is located at a significantly smaller velocity of about 24~\kms. These systematic velocity shifts between Stokes I and V can have different origins. A first possible explanation would involve the Stokes V asymmetry reported above, as it can induce a shift of the zero-crossing in the case of a limited spectral resolution (Solanki \& Stenflo 1986); but in this case, the sharp evolution of $da$ and $dA$ should progressively displace the zero-crossing towards lower radial velocities, which is not observed. A second option involves a large-scale toroidal magnetic field\index{magnetic field} (Petit et al. 2005), but the existence of a toroidal component would be hard to reconcile with the absence of any significant rotation in Betelgeuse\index{Betelgeuse} (unless the slowly rotating surface is hiding faster rotation in internal layers). As a third option, we propose that the red-shift of Stokes V profiles may be related to the concentration of magnetic elements in the sinking component of the convective mixing, as observed in small-scale solar magnetic elements. If this last interpretation seems more easy to reconcile with the basic properties of Betelgeuse\index{Betelgeuse} and is also quantitatively consistent with the convective velocity measurements of Gray (2008), further investigation is obviously needed before reaching any conclusion.

\section{A small-scale dynamo\index{dynamo} in Betelgeuse\index{Betelgeuse} ?}

In the past, strong magnetic fields\index{magnetic field} have been detected in fast-rotating giants or sub-giants belonging to RS CVn systems or to the FK Com class (Petit et al. 2004a, 2004b). More recently, repeated magnetic field\index{magnetic field} detections have been obtained in active, single red giants (Konstantinova et al. 2008, 2010, Auri\`ere et al. 2009, Sennhauser \& Berdyugina 2011) and in a likely descendant of a strongly magnetic Ap star\index{star} (Auri\`ere et al. 2008). 

In this growing literature on evolved stars\index{star}, the detection of a weak surface magnetic field\index{magnetic field} at the surface of Betelgeuse\index{Betelgeuse} (Auri\`ere et al. 2010) is an important observational result, in the sense that the physical interpretations proposed for other objects to account for their magnetic nature cannot be applied here. Firstly, the magnetic field\index{magnetic field} of Betelgeuse\index{Betelgeuse} has to be generated without the help of a fast, or even moderate stellar rotation, and this specificity should exclude any global dynamo\index{dynamo}. Secondly, the very large radius implies that any magnetic remnant of a strong magnetic field\index{magnetic field} on the main sequence would be too diluted to be detectable at photospheric level. In this situation, a more natural interpretation would involve the convection alone as the engine of a dynamo\index{dynamo}, bringing the first strong observational evidence that such process (proposed by Dorch \& Freytag 2003) can be efficient in cool stars\index{star}.

This exciting result, later confirmed for a larger sample of cool supergiants\index{cool supergiant} (Grunhut et al. 2010), comes together with a number of additional tracers of magnetic activity\index{activity} and convection (chromospheric emission, radial velocities, line bisectors, Stokes V asymmetries). This wealth of information is a motivation to pursue the spectropolarimetric\index{spectropolarimetry} monitoring of Betelgeuse\index{Betelgeuse}, in order to investigate longer-term trends that may affect the various measurements at our disposal and study the possible role of the surface magnetic field\index{magnetic field} in the onset of the mass-loss of Betelgeuse\index{Betelgeuse} and other supergiant stars\index{star}.

\end{document}